\documentclass[12pt]{article}
\usepackage{latexsym}
\usepackage[latin1]{inputenc}
\usepackage[T1]{fontenc}
\title{Four--dimensional Gravity from Singular Spaces}
\author{Ph.~Brax \\
{\it Service de Physique Th\'eorique}\\
{\it CEA-Saclay}\\
{\it F-91191, Gif/Yvette cedex, France}\\
{\it }\\
C.~van~de~Bruck, A.~C.~Davis, C.~S.~Rhodes \\
{\it Department of Applied Mathematics and Theoretical Physics}\\
{\it Centre for Mathematical Sciences}\\
{\it Wilberforce Road, University of Cambridge}\\
{\it CB3 0WA, United Kingdom}}
\begin{document}
\maketitle

\begin{abstract} 
  The modification to four--dimensional Einstein gravity at low 
  energy in two brane models is investigated 
  within supergravity in singular spaces. 
  Using perturbation theory around a static BPS background, we study
  the effective four--dimensional gravitational theory, a
  scalar--tensor theory, and derive the Brans--Dicke parameter when
  matter is present on the positive tension brane only. We show there
  is an attractor mechanism towards general relativity in the matter
  dominated era. The dynamics of the interbrane distance are discussed.
  Finally, when matter lives on both branes, we find that there is a
  violation of the equivalence principle whose magnitude is governed
  by the warping of the extra dimension.
\end{abstract}

\vspace{0.5cm}

\noindent {\small DAMTP-2002-25, t02/009}

\section{Introduction}

The notion that our Universe may be a hypersurface within a
higher-dimensional space-time is not a new one
\cite{akam1,ruba,visser}, but recently interest in this class of
theory of the Universe has been rekindled by the development of simple
models, notably by Randall and Sundrum \cite{rshierarchy99} and
\cite{rscompact99}.  These models admit a rich set of developments and
extensions. In brane world theories the standard model fields are
usually confined on our (visible) brane, while allowing gravity to
propagate throughout the entire space-time.  Recovering a
four-dimensional theory of gravity that is compatible with present-day
observations is one of the main constraints to be imposed on
brane-world models. Such an analysis has been carried out in the
RS framework with an explicit radion stabilization mechanism in \cite{Ta}.

One possible extension to the RS set-up is the inclusion of matter
fields in the (extra-dimensional) bulk. In particular it is of
interest to study the dynamics of bulk scalar fields as these fields
appear naturally in models motivated from M--theory \cite{Lukas}, see
\cite{BD1}--\cite{Himemoto}.
Recently the issue of retrieving gravity at low energy for models with a single
bulk scalar field has been analyzed in \cite{kof}. It has been found that
four dimensional gravity is generically recovered provided that the models
do not have a super-gravitational origin\footnote{More specifically,
the authors of \cite{kof} assume that the relation $U=U_B'^2-U_B^2$ between
the brane and the bulk potentials is violated. This relation holds explicitly
for the models that we consider.}.

In this paper, we discuss the effect of including a scalar field in
the bulk, for certain super-gravitational brane-world models
\cite{BD1,BD2,BD3,BD4}, on the effective four-dimensional theory of
gravity. To examine this, we build on the results of a companion paper
\cite{Ef} in which we studied the dynamics of linear perturbations of
the radion (i.e. fluctuations of the brane distance) and the bulk
scalar field.  We compute the metric perturbation due to matter on the
brane, and compare the low-energy limit to the Brans--Dicke theories;
see \cite{Chiba}. 

We begin by summarizing the relevant results from our companion paper
in section \ref{sec:radion} and add matter sources on the branes; we
then, in section \ref{sec:grav}, compute the Brans--Dicke parameter
that would be observed on the positive-tension brane in the low-energy
limit, and show that general relativity is an attractor in rather general
cases. We draw our conclusions in section \ref{sec:conc}.

\section{Brane World Setup: Background and Perturbations}
\label{sec:radion}
\subsection{The Background Configurations}
We are interested in brane world models with a bulk scalar field. 
In addition, we have two branes in our setup: one (the visible brane)
is located at $z=0$, whereas the other (hidden) brane is located at 
$z=r_c$. We consider a bulk action
\begin{equation}
  S_{\mbox{\scriptsize bulk}} =
  \frac{1}{2\kappa_5^2}\int d^5 x \sqrt{-g_5}\left(R-\frac{3}{4}((\partial \phi)^2 +U)\right)
\end{equation}
with boundary terms
\begin{equation}
  S_{\mbox{\scriptsize brane}}=-\frac{3}{2\kappa_5^2}\int d^5 x\sqrt {-g_5}(\delta
  (z)-\delta (z-r_c))U_B
\end{equation}
where $U_B$ is the superpotential related to the bulk potential $U$ by
\begin{equation}
  U=\left(\frac{\partial U_B}{\partial \phi}\right)^2-U_B^2.
\end{equation}

Background solutions to the equations of motion are described by
the BPS equations
\begin{equation}
  \label{eq:BPS}
  \frac{a'}{a}=-\frac{U_B}{4},\ \phi'=\frac{\partial U_B}{\partial \phi}
\end{equation}
where $'=d/dz$, for a metric of the form
\begin{equation}
  ds^2=dz^2+ a^2(z)\eta_{\mu\nu}dx^\mu dx^\nu
\end{equation}
We will particularly focus on the case where
\begin{equation}
  U_B=4k e^{\alpha \phi}
\end{equation}
with $\alpha =1/\sqrt 3,-1/\sqrt {12}$; these values correspond to
supergravity in singular spaces \cite{BD1}.  In those cases the
solution reads
\begin{equation}
  a(z)=\left(\frac{1-4k\alpha^2(z-z_0)}{1+4k\alpha^2z_0}\right)^{\frac{1}{4\alpha^2}}
\end{equation}
where the positive tension brane sits at $z=0$.
In the $\alpha\to 0$ limit we retrieve the AdS profile
\begin{equation}
  a(z)=e^{-kz}
\end{equation}
Notice that in that case the scalar field decouples altogether, as
$U_B$ and hence $U$ are constant.

\subsection{Fluctuations}

Our aim is to analyse the influence of the
matter localized on the brane. We thus include matter on the 
boundary branes which will be treated as {\it perturbations}. We
will further assume that the matter does not couple to the bulk scalar 
field, so that, in particular, the junction conditions for the bulk 
field are not modified. As a consequence we are able to build on the
analysis performed in \cite{Ef}.

We consider a perturbed bulk metric
\begin{equation}
ds^2= dz^2 + a^2(z)dx^{\mu}dx_{\mu}+ h_{\mu\nu}dx^{\mu}dx^{\nu}
\end{equation}
in Gaussian normal coordinates. The dynamics of the scalar field
fluctuation is determined by the Klein-Gordon equation  
\begin{equation}
\Box \delta\phi+ \frac{1}{2}h'\phi'= \frac{1}{2}\frac{\partial^2 U_B}{\partial
\phi^2} \delta \phi 
\end{equation}
where $'=d/dz$ and $h$ is the trace of $h_{\mu\nu}$. 
The zero-mode solution to this equation can be obtained as
\begin{equation}
\delta\phi= \frac{\partial U_B}{\partial \phi} \hat \phi(x^{\mu})
\end{equation}
and $h=0$, which satisfies the boundary conditions at both branes for any
massless $\hat\phi(x^\mu)$.
 
Similarly one can write the Einstein equations in the bulk 
\begin{equation}
\Box h_{ab}+\frac{U_B^2}{8}h_{ab}- D_{\{a}D_ch^c_{b\}}=\frac{3}{4}\partial_{\{a}\delta\phi\partial_{b\}}\phi_0
\end{equation}
where we have extended $h_{\mu\nu}$ to five dimensions with
$h_{55}=h_{\mu5}=0$. 
The $(55)$ component of the gravitational equation
leads to
$\partial_z \delta\phi=0$.
This implies that $\delta{\phi}$ is a function of $x^{\mu}$ only and therefore
\begin{equation}
\delta\phi=0 
\end{equation}
So the scalar field perturbation vanishes altogether.
Now the $(\mu 5)$ equation leads to
\begin{equation}
D_{\mu}h^{\mu\nu}=0
\end{equation}
In summary, we find that the scalar field does not play a r\^{o}le in
the low energy dynamics and that the metric perturbation is transverse
and traceless.

\subsubsection{Adding matter sources on the branes}

The effect of adding matter on the branes is to induce a bending of
the brane. This leads to a shift in the position of both branes which
sit at
\begin{equation}
z_+=\xi^+,\ z_-=r_c + \xi^-
\end{equation}
One can perform local changes of coordinates around both branes which 
preserve the Gaussian normal coordinates and shift them to the origin 
and $r_c$. After such changes of coordinates the metric is
\begin{equation}
h_{\mu\nu}^+= h_{\mu\nu}
-\frac{U_B}{2}a^2(z)\xi^+\eta_{\mu\nu}-2a^2(z)\int_0^z
\frac{dy}{a^2(y)}\partial_{\mu}\partial_{\nu}\xi^+
\end{equation}
in the vicinity of the origin and 
\begin{equation}
h_{\mu\nu}^-= h_{\mu\nu}
-\frac{U_B}{2}a^2(z)\xi^-\eta_{\mu\nu}-2a^2(z)\int_{r_c}^z
\frac{dy}{a^2(y)}\partial_{\mu}\partial_{\nu}\xi^-
\end{equation}
in the vicinity of the second brane.

We can now study these perturbations. 
To do so we come back to the gravitational equation; it is convenient
to work in the conformal gauge, defining 
\begin{equation}
du = \frac{dz}{a}
\end{equation}
and
\begin{equation}
  h_{\mu\nu}=\sqrt  a \tilde h_{\mu\nu}.
\end{equation}
We denote by $u_+$ and $u_-$ the location of both branes in the
conformal coordinate system. 
The gravitational equations then read
\begin{equation}
  -\frac{d^2\tilde h_{\mu\nu}}{du^2}+ \frac{1}{a^{3/2}}\frac{d^2
    a^{3/2}}{du^2}\tilde h_{\mu\nu}=\Box^{(4)}\tilde h_{\mu\nu}
  \label{eq:gra}
\end{equation}
or
\begin{equation}
  \left[Q^\dagger Q - \Box^{(4)}\right]\tilde{h}_{\mu\nu}=0,
\end{equation}
with
\begin{equation}
  Q=-\frac{d}{du}+\frac{d\ln a^{3/2}}{du},\
  Q^{\dagger}=\frac{d}{du}+\frac{d\ln a^{3/2}}{du},\
  \Box^{(4)}=\eta^{\mu\nu}\partial_{\mu}\partial_{\nu}.
\end{equation}
The boundary conditions are now 
\begin{equation}
  \left.\partial_z h_{\mu\nu}\right|_{u_{\pm}}+\left.\frac{U_B}{2}h_{\mu\nu}\right|_{u_{\pm}}=-\kappa_5^2\Sigma^{\pm}_{\mu\nu}
\end{equation}
where
\begin{equation}
  \Sigma^{\pm}_{\mu\nu}=
  -\frac{2}{\kappa_5^2}\partial_{\mu}\partial_{\nu}\xi^{\pm}
  +(T_{\mu\nu}^{\pm}-\frac{1}{3}T^{\pm}g_{\mu\nu}^{\pm})
\end{equation}
and $T^{\pm}_{\mu\nu}$ is the matter energy-momentum tensor on both
branes while $g_{\mu\nu}^{\pm}$ is the induced metric. 

The equations of motion can be conveniently written as
\begin{equation}
  Q^{\dagger}Q h_{\mu\nu}-\Box^{(4)}\tilde h_{\mu\nu}=
\sum_{\sigma=\pm}  \left(\frac{3}{4}U_B h^{\sigma}_{\mu\nu}+2\kappa_5^2\sqrt a \Sigma^{\sigma}_{\mu\nu}\right)\delta(u-u_{\sigma})
  \end{equation}
The brane equation of motion determining the brane bending is obtained
by taking the trace of the previous equation.
Using the tracelessness of $h_{\mu\nu}$ we conclude that 
\begin{equation}
\Box^{(4)}\xi^{\pm}=-\frac{\kappa_5^2}{6}a_{\pm}^2 T^{\pm}
\end{equation}
where $a_{\pm}=a(u_{\pm})$.

Let us decompose into Kaluza-Klein modes
\begin{equation}
  \Box^{(4)} h_{\mu\nu}=-k^2h_{\mu\nu}
\end{equation}
It is convenient to define the eigenstates of the 
positive operator $Q^{\dagger}Q$
\begin{equation}
  Q^{\dagger}Q\psi_{\lambda}=\lambda^2 \psi_{\lambda}
\end{equation}
Consider now the Green's function $G_{k^2}$ of the operator
\begin{equation}
  H_{k^2}=Q^{\dagger}Q+k^2-\frac{3}{4}U_B(\delta(u-u_+)+\delta(u-u_-))
\end{equation}
defined by
\begin{equation}
  G_{k^2}(u,u')=\sum_{\lambda}\frac{\psi_{\lambda}(u)\psi_{\lambda}(u')}{\lambda^2
 + k^2 +i\epsilon}
\end{equation}
where each mode is normalized.  The only  mode with $\lambda=0$ is
proportional to $a^{\frac{3}{2}}$, as it is the only zero mode  to satisfy the
boundary conditions imposed by the delta functions. Moreover the spectrum is
discrete due to the two boundary conditions.

We can now write the metric perturbation due to matter on both branes 
\begin{equation}
   h_{\mu\nu}(u,x)=2\kappa_5^2\sum_{\sigma=\pm}\int d^4x'
G(x,u;x',u_{\sigma}) 
a_{\sigma}^{1/2}a^{1/2}(u)\Sigma^\sigma _{\mu\nu}(x')
\end{equation}
where
\begin{equation}
  G(x,u;x',u')=\int \frac{d^4k}{(2\pi)^4} e^{ik.(x-x')}G_{k^2}(u,u')
\end{equation}
This provides the solution to the Einstein equations in the bulk due
to the brane bending. 

\section{Gravity on the visible brane: the zero mode truncation}
\label{sec:grav}

We concentrate first on the case where no matter is present on the second
brane and study gravity on the first brane. 
In the low energy limit we neglect the contribution due to the
Kaluza-Klein modes and retain the zero mode only.
In this case the metric perturbation on the brane is
\begin{equation}
  h_{\mu\nu}=-\frac{\kappa_5^2}{\int_0^{r_c} dz  a^2}\frac{1}{\Box^{(4)}}(T^+_{\mu\nu}-\frac{1}{3}T^+\eta_{\mu\nu})
  -\left.\frac{U_B}{2}\right|_{0} \xi^+\eta_{\mu\nu},
\end{equation}
where we have normalized $a_+=1$ and used a four dimensional gauge
transformation to get rid of the second derivative terms. 
This leads to 
\begin{equation}
  \Box^{(4)}{h_{\mu\nu}}=-\frac{\kappa_5^2}{\int_0^{r_c} dz  a^2}(T^+_{\mu\nu}-\frac{1}{3}T^+\eta_{\mu\nu})
  +\frac{U_B}{12}\vert _0 \kappa_5^2 T^+\eta_{\mu\nu},
\end{equation}
Corresponding to the Brans--Dicke expression
\begin{equation}
  \Box^{(4)}h_{\mu\nu}=-\frac{16\pi G_N}{\Phi}(T^+_{\mu\nu}
  -\frac{1}{2}T^+\eta_{\mu\nu})
  -\frac{8\pi G_N}{\Phi (3+2\omega(\Phi))} T^+\eta_{\mu\nu},
\end{equation}
where $\Phi$ is the Brans--Dicke field and $\omega(\Phi)$ the
Brans--Dicke parameter.
We identify
\begin{equation}
  \kappa_5^2=16\pi G_N \int^{r_{\infty}}_0 \frac{a^2}{a_+^2} dz, 
\end{equation}
where $r_{\infty}$ is the maximal range of the extra dimension,
i.e.\ infinity for the Randall-Sundrum scenario and the singularity when a
bulk scalar field is present.
Next we identify
\begin{equation}
  \Phi=\frac{\int^{r_c}_0a^2 dz}{\int^{r_{\infty}}_0 a^2 dz},
\end{equation}
which varies between zero and one.
Finally we obtain the Brans--Dicke parameter
\begin{equation}
  \omega(\Phi)= \frac{3}{2}\left(-1+\frac{1}{1-\left.\Phi\frac{U_B}{2}\right|_0
\int_0^{r_{\infty}}\frac{a^2}{a_+^2} dz}\right)
\end{equation}
We can now consider the specific exponential models, with
$U_B=4ke^{\alpha\phi}$.
This leads to
\begin{equation}
  \Phi=1-\left(1-\frac{r_c}{r_{\infty}}\right)^{1+\frac{1}{2\alpha^2}},
\end{equation}
where $r_{\infty}=z_0+1/4k\alpha^2$.
Similarly for the Brans--Dicke parameter
\begin{equation}
\omega(\Phi)=\frac{3}{2}\left( - 1+\frac{1}{1- \Phi
\frac{1+4k\alpha^2z_0}{1+2\alpha^2}} \right).
\end{equation}
In the $\alpha\to 0$ limit we retrieve that
\begin{equation}
  \Phi=1-e^{-2kr_c}
\end{equation}
and
\begin{equation}
  \omega(\Phi)=\frac{3}{2}\left(-1+e^{2kr_c}\right),
\end{equation}
in agreement with \cite{gravrs99}.  

Let us finally comment on the well known attractor--mechanism for 
scalar tensor theories in the context of the theory discussed so 
far. Damour and Nordtvedt \cite{Damour} found that in the case for Brans--Dicke 
theory the dynamics of the scalar field evolves during the matter 
dominated epoch such that the Brans--Dicke parameter grows rapidly 
and thus that Einstein gravity is a cosmological attractor. 
It is easy to verify that this is the case in the theory above:

First, it is convenient to go to the Einstein frame where 
$g_{\mu\nu}=g_{E,\mu\nu}/\Phi$. We then define a canonically normalized 
field $\sigma$ by $\partial a/\partial \sigma= \beta(\sigma)$ where 
$a(\sigma)= -\ln (\Phi)/2$ and $\beta^2(\sigma)=1/(3+2\omega(\Phi))$, 
and find that
\begin{equation}
  \beta^2(\sigma)=1-\frac{\Phi}{\Phi_0},
\end{equation}
where $\Phi_0=(1+2\alpha^2)/(1+4\alpha^2 kz_0)$.
This leads to
\begin{equation}
  \sigma-\sigma_0=\frac{1}{2}\ln \left(\frac{1+\beta}{1-\beta}\right).
\end{equation}
Notice that $\beta \to 0$ and $\sigma \to \sigma_0$ as $\Phi\to \Phi_0$.
The field $\sigma$ is attracted towards value its value at $\beta=0$,
i.e. $\sigma$ is attracted towards $\sigma_0$. The minimum
$\sigma=\sigma_0$ can only been reached when $\Phi_0\le 1$, i.e.
for $z_0\ge 1/2k$. In that case the field $\sigma$ is attracted towards
the point where $\omega(\Phi)=+\infty$, allowing us to recover
Einstein gravity.

In the opposite situation ($\Phi_0 > 1$) the Brans--Dicke parameter can
only reach a maximal value
\begin{equation}
  \omega_{\mbox{\scriptsize max}} =
  \frac{3}{2}\frac{1+4\alpha^2kz_0}{2\alpha^2\left(1-2kz_0\right)}
\end{equation}
which, since $\Phi_0 > 1$ implies that $2kz_0 < 1$, is in general not
large enough to escape the experimental bounds on the Brans--Dicke
parameter (and tends to $\frac{3}{4\alpha^2}$ in the limit as $2kz_0
\to 0$).

In conclusion, we have found that the Brans--Dicke parameter converges
towards infinity as soon as the distances between the branes and the
naked singularity are large enough. This should be the case for a 
realistic cosmology, because we expect both branes to expand. Thus, 
they will move away from the singularity so that the attractor
mechanism can operate. 

Let us finally discuss the case where matter is also on the second
brane and study its effect on the gravitational field as seen on the 
first brane. It is to be expected that the leading effect will be due 
to the propagation of gravitons in the bulk, hence suppressed by the 
warping of the bulk. Following the same method as in the previous 
calculation we find that
\begin{equation}
\frac{1}{a_+^2}\Box^{(4)}h_{\mu\nu}= -\frac{16\pi
G_N}{\Phi}(T_{\mu\nu}
-\frac{1}{3} (T_{\rho\lambda}\eta^{\rho\lambda})\eta_{\mu\nu})+
\frac{U_B}{12}\vert_0 (T^+_{\rho\lambda}\eta^{\rho\lambda})\kappa_5^2 \eta_{\mu\nu},
\end{equation}
where the total energy momentum tensor is
\begin{equation}
T_{\mu\nu}=T^+_{\mu\nu}+\frac{a_-^2}{a_+^2}T_{\mu\nu}^-.
\end{equation}
This can be recast into the form
\begin{eqnarray} 
\frac{1}{a_+^2}\Box^{(4)}h_{\mu\nu} &=& -\frac{16\pi
G_N}{\Phi}(T_{\mu\nu}
-\frac{1}{2} (T_{\rho\lambda}\eta^{\rho\lambda})\eta_{\mu\nu})
-\frac{8\pi G_N}{\Phi(3+2\omega(\Phi))}
(T_{\rho\lambda}\eta^{\rho\lambda})\eta_{\mu\nu} \nonumber \\
&-&\frac{k}{3}\frac{a_-^2}{a_+^2}\kappa_5^2 (\eta^{\rho\lambda}T^-_{\rho\lambda})\eta_{\mu\nu},
\end{eqnarray}
where $G_N$, $\Phi$ and $\omega(\Phi)$ were defined above. 
Notice that there is a direct coupling to matter on the second brane
which does not follow from Brans--Dicke theory.
In particular this implies that gravity does not couple universally 
to matter, i.e this leads to a violation of the equivalence principle
whose strength is proportional to the warping factor $a_-^2/a_+^2$.

\section{Conclusions}
\label{sec:conc}
We have presented an analysis of the effective four-dimensional
gravity at the positive-tension brane for a brane-world model with
bulk scalar field. When the BPS conditions (\ref{eq:BPS}) are imposed on
the system, generating a static supergravity background, we find that
the effective low-energy theory is a Brans--Dicke theory. We have briefly 
discussed the attractor mechanism which attracts the theory towards 
general relativity during a matter dominated epoch. We have found that the 
conditions for this to happen can easily be fulfilled, as long as 
the singularity is far away from both branes. 

When matter is present on both branes, we have found that both matter types 
couple differently to gravity. Thus, we expect a violation of the 
equivalence principle for matter which lives on the shadow brane. 
This violation, however, is suppressed by a factor, which can be 
very small, when the branes are far apart. We expect very interesting 
consequences for cosmology from these considerations, which will be
investigated in future work. 

\paragraph*{Acknowledgements}

This work was supported by PPARC (A.C.D., C.S.R. and C.v.d.B.), 
Jesus College, Cambridge (C.S.R.), the Deutsche 
Forschungsgemeinschaft (C.v.d.B.), a CNRS--Royal Society exchange 
grant for collaborative research and the European network (RTN), 
HPRN-CT-2000-00148 and PRN-CT-2000-00148.


\begin{thebibliography}{10}

\bibitem{akam1}
K.~Akama, {\it {An Early Proposal of ``Brane World''}},  {\em Lect.\ Notes.\
  Phys} {\bf 176} (1982) 267--271,
  [\href{http://xxx.lanl.gov/abs/hep-th/0001113}{{\tt hep-th/0001113}}].

\bibitem{ruba}
V.~A. Rubakov and M.~E. Shaposnikov, {\it {Do we live inside a domain wall?}},
  {\em Phys.\ Lett.} {\bf B125} (1983) 136.

\bibitem{visser}
M.~Visser, {\it {An exotic class of Kaluza-Klein models}},  {\em Phys.\ Lett.}
  {\bf B159} (1985) 22, [\href{http://xxx.lanl.gov/abs/hep-th/9910093}{{\tt
  hep-th/9910093}}].

\bibitem{rshierarchy99}
L.~Randall and R.~Sundrum, {\it {A Large Mass Hierarchy from a Small Extra
  Dimension}},  {\em Phys.\ Rev.\ Lett.} {\bf 83} (1999) 3370--3373,
  [\href{http://xxx.lanl.gov/abs/hep-th/9905221}{{\tt hep-th/9905221}}].

\bibitem{rscompact99}
L.~Randall and R.~Sundrum, {\it {An Alternative to Compactification}},  {\em
  Phys.\ Rev.\ Lett.} {\bf 83} (1999) 4690--4693,
  [\href{http://xxx.lanl.gov/abs/hep-th/9906064}{{\tt hep-th/9906064}}].

\bibitem{Ta} X. Montes and T. Tanaka, {\it {Gravity in the Brane World
for Two--Branes Model with Stabilized Modulus}}, 
{\em Nucl. Phys.} {\bf B582} (2000) 259,
[\href{http://xxx.lanl.gov/abs/hep-th/9906064}{{\tt hep-th/0001092}}].

\bibitem{Lukas}
A.~Lukas, B.~Ovrut, K.~S. Stelle, and D.~Waldram, {\it {The Universe as a
  Domain Wall}},  {\em Phys.\ Rev.} {\bf D59} (1999) 086001,
  [\href{http://xxx.lanl.gov/abs/hep-th/9803235}{{\tt hep-th/9803235}}].

\bibitem{BD1}
P.~Brax and A.-C. Davis, {\it {Cosmological Solutions of Supergravity in
  Singular Spaces}},  {\em Phys.\ Lett.} {\bf B497} (2001) 289--295,
  [\href{http://xxx.lanl.gov/abs/hep-th/0011045}{{\tt hep-th/0011045}}].

\bibitem{BD2}
P.~Brax and A.-C. Davis, {\it {Cosmological Evolution on Self-Tuned Branes and
  the Cosmological Constant}},  {\em \rm JHEP} {\bf 0105} (2001) 008,
  [\href{http://xxx.lanl.gov/abs/hep-th/0104023}{{\tt hep-th/0104023}}].

\bibitem{BD3}
P.~Brax, C.~van~de Bruck, and A.-C. Davis, {\it {Brane-World Cosmology, Bulk
  Scalars and Perturbations}},  {\em \rm JHEP} {\bf 0110} (2001) 026,
  [\href{http://xxx.lanl.gov/abs/hep-th/0108215}{{\tt hep-th/0108215}}].

\bibitem{BD4}
P.~Brax and A.-C. Davis, {\it {On Brane Cosmology and Naked Singularities}},
  {\em Phys.\ Lett.} {\bf B513} (2001) 156--162,
  [\href{http://xxx.lanl.gov/abs/ep-th/0105269}{{\tt hep-th/0105269}}].

\bibitem{Ef}
P.~Brax, C.~van~de Bruck, A.-C. Davis, and C.~S. Rhodes, {\it {Wave function
  of the radion with bulk scalar field}}, to appear in Phys. Lett. B.,
  \href{http://xxx.lanl.gov/abs/hep-th/0201191}{{\tt hep-th/0201191}}.

\bibitem{Langlois}
D.~Langlois and M.~Rodriguez-Martinez, {\it {Brane cosmology with a bulk scalar
  field}},  {\em Phys.\ Rev.} {\bf D64} (2001) 123507,
  [\href{http://xxx.lanl.gov/abs/hep-th/0106245}{{\tt hep-th/0106245}}].

\bibitem{Mennim}
A.~Mennim and R.~Battye, {\it {Cosmological expansion on a dilatonic
  brane-world}},  {\em Class.\ Quant.\ Grav.} {\bf 18} (2001) 2171--2194,
  [\href{http://xxx.lanl.gov/abs/hep-th/0008192}{{\tt hep-th/0008192}}].

\bibitem{Wands}
K.~Maeda and D.~Wands, {\it {Dilaton-gravity on the brane}},  {\em Phys.\ Rev.}
  {\bf D62} (2000) 124009, [\href{http://xxx.lanl.gov/abs/hep-th/0008188}{{\tt
  hep-th/0008188}}].

\bibitem{Kunze}
A.~Feinstein, K.~E. Kunze, and M.~A. Vazquez-Mozo, {\it {Curved dilatonic brane
  worlds}},  {\em Phys.\ Rev.} {\bf D64} (2001) 084015,
  [\href{http://xxx.lanl.gov/abs/hep-th/0105182}{{\tt hep-th/0105182}}].

\bibitem{Flanagan}
E.~E. Flanagan, S.-H.~H. Tye, and I.~Wasserman, {\it {Brane World Models With
  Bulk Scalar Fields}},  {\em Phys.\ Lett.} (2001) 155,
  [\href{http://xxx.lanl.gov/abs/hep-th/0110070}{{\tt hep-th/0110070}}].

\bibitem{Davis}
S.~C. Davis, {\it {Brane Cosmology Solutions with Bulk Scalar Fields}},
  \href{http://xxx.lanl.gov/abs/hep-ph/0111351}{{\tt hep-ph/0111351}}.

\bibitem{csaki2}
C.~Csáki, J.~Erlich, C.~Grojean, and T.~J. Hollowood, {\it {General Properties
  of the Self-tuning Domain Wall Approach to the Cosmological Constant
  Problem}},  {\em Nucl.\ Phys.} {\bf B584} (2000) 359--386,
  [\href{http://xxx.lanl.gov/abs/hep-th/0004133}{{\tt hep-th/0004133}}].

\bibitem{cline}
J.~Cline, J.~Vinet, {\it {Order $\rho^2$ Corrections to
Randall--Sundrum I Cosmology}}, 
\href{http://xxx.lanl.gov/abs/gr-qc/0112027}{{\tt gr-qc/0201041}}.


\bibitem{Himemoto}
Y.~Himemoto, T.~Tanaka, and M.~Sasaki, {\it {A bulk scalar in the braneworld
  can mimic the 4d inflaton dynamics}},
  \href{http://xxx.lanl.gov/abs/gr-qc/0112027}{{\tt gr-qc/0112027}}.

\bibitem{kof} 
S.~Mukohyama and L.~Kofman, {\it {Brane Gravity at Low Energy}}, 
[\href{http://xxx.lanl.gov/abs/hep-th/9906064}{{\tt hep-th/0112115}}].

\bibitem{Chiba}
T.~Chiba, {\it {Scalar-Tensor Gravity in Two 3-brane System}},  {\em Phys.\
  Rev.} {\bf D62} (2000) 021502,
  [\href{http://xxx.lanl.gov/abs/gr-qc/0001029}{{\tt gr-qc/0001029}}].

\bibitem{gravrs99}
J.~Garriga and T.~Tanaka, {\it {Gravity in the Randall-Sundrum Brane World}},
  {\em Phys.\ Rev.\ Lett.} {\bf 84} (2000) 2778--2781,
  [\href{http://xxx.lanl.gov/abs/hep-th/9911055}{{\tt hep-th/9911055}}].

\bibitem{Damour}
T.~Damour and K.~Nordtvedt, {\it {Tensor-scalar cosmological models and their
  relaxation toward general relativity}},  {\em Phys.\ Rev.} {\bf D48} (1993)
  3436.

\end{thebibliography}

\providecommand{\href}[2]{#2}\begingroup\raggedright\endgroup

\end{document}